\documentclass[reprint,aps,prr,superscriptaddress,amsmath,amssymb,floatfix]{revtex4-2}
\usepackage{graphicx}
\usepackage{layout}
\usepackage{natbib}
\usepackage{dcolumn}
\usepackage{braket}
\usepackage{bm}
\usepackage{color}
\usepackage{verbatim}
\usepackage{hyperref}
\usepackage{soul}
\usepackage[textsize=tiny]{todonotes}

\setstcolor{red}

\begin{document}

\title{Synthetic fractional flux quanta in a ring of superconducting  qubits}

\author{Luca Chirolli}
\affiliation{Quantum Research Center, Technology Innovation Institute, Abu Dhabi, UAE}

\author{Juan Polo}
\affiliation{Quantum Research Center, Technology Innovation Institute, Abu Dhabi, UAE}

\author{Gianluigi Catelani}
\affiliation{Quantum Research Center, Technology Innovation Institute, Abu Dhabi, UAE}
\affiliation{JARA Institute for Quantum Information (PGI-11),Forschungszentrum J\"ulich, 52425 J\"ulich, Germany}

\author{Luigi Amico}
\affiliation{Quantum Research Center, Technology Innovation Institute, Abu Dhabi, UAE}
\affiliation{Dipartimento di Fisica e Astronomia 'Ettore Majorana', Via S. Sofia 64, 95123 Catania, Italy}
\affiliation{INFN-Sezione di Catania, Via S. Sofia 64, 95127 Catania, Italy}

\begin{abstract}
A ring of capacitively coupled transmons threaded by a synthetic magnetic field is studied as a realization of a
strongly interacting bosonic system. The synthetic flux is imparted through a specific Floquet modulation scheme
based on a suitable periodic sequence of Lorentzian pulses that are known as “Levitons.” Such scheme has the
advantage to preserve the translation invariance of the system and to work at the qubit sweet spots. We employ
this system to demonstrate the concept of fractional values of flux quanta. Although such fractionalization
phenomenon was originally predicted for bright solitons in cold atoms, it may be in fact challenging to access
with that platform. Here, we show how fractional flux quanta can be read out in the absorption spectrum of a
suitable “scattering experiment” in which the qubit ring is driven by microwaves.
\end{abstract}

\maketitle

\section{Introduction}
Networks of mesoscopic-scale systems at sufficiently low temperature can provide us with artificial quantum matter that can be exploited both to study fundamental aspects of quantum science and to design new quantum devices~\cite{imry2002intro}. To this end, different platforms have been explored, ranging from arrays of quantum dots~\cite{michler2017quantum} and trapped ions~\cite{duan2010colloquium} to superconducting circuits~\cite{kjaergaard2020superconducting} and cold atoms~\cite{bloch2012quantum,bluvstein2021controlling}. A fruitful viewpoint adopted so far in the  community has been that genuine notions of mesoscopic physics of electronic systems such as the Josephson effect, point contacts, or persistent currents, can provide inspiration to define new concepts in quantum technology~\cite{krinner2014observation,valtolina2015josephson,singh2024shapiro,amico2022colloquium,Amico2021roadmap,polo2023perspective}. Here, we provide a case study  going in the opposite direction: We demonstrate that specific notions originally emerging in current research in quantum technology, as in ultracold atoms systems, can inspire electronic mesoscopic physics. Specifically, we will refer to attractive bosonic cold atoms~~\cite{carr2000stationary,strecker2002formation,khaykovich2002formation,nguyen2017formation,marchant2013controlled}. On one hand, such  systems  are difficult to simulate classically. On the other hand,  attracting bosons allow the formation of bright solitons that, in turn, bear a great potential in atomtronics and quantum sensing~\cite{amico2022colloquium,Amico2021roadmap,polo2023perspective}. In particular, bright solitons in a one-dimensional lattice have peculiar properties of stability~\cite{naldesi2019rise}, which are predicted to have a strong impact in interferometry~\cite{naldesi2023massive}. 

Here, we focus on a striking property of lattice  bright solitons confined in a ring-shaped potential pierced by a synthetic magnetic field: As a result of specific  many-body correlations, the system's persistent current is characterized by a periodicity reflecting a fractional value with respect to the flux quantum $\Phi_0$ obtained in the non-interacting case. 
Crucially for the logic we adopt in the present work, we note that the fractionalization phenomenon manifests itself in a fractional periodicity of the energies of the system as function of the magnetic field, which, following Leggett, are the analogous of Bloch bands for the problem~\cite{leggett1991theorem}. In the present  case, such fractionalization is predicted to scale as the inverse of the number of particles $N_p$ in the system~\cite{naldesi2022enhancing}. Such $N_p$ dependence makes it difficult to directly observe the feature in  cold atoms experiments, since  $N_p$ is typically large.

\begin{figure}[t]
	\centering
    \includegraphics[width=1\linewidth]{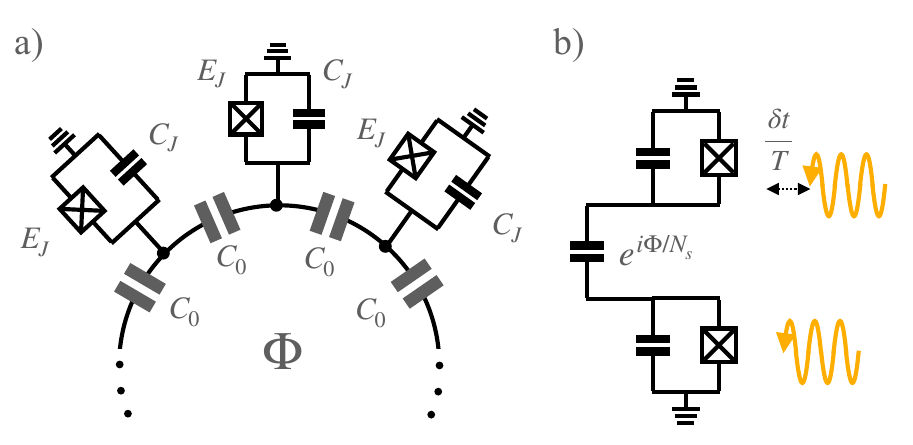}
	\caption{(a) Schematics of a circuit constituted by a chain of transmons, which realizes an attractive Bose-Hubbard model. (b) Floquet modulation protocol, consisting in local transmon frequency modulation. }
	\label{Fig1}
\end{figure}

In this work, we will define a superconducting qubit quantum simulator for the dynamics of attracting bosons hosted in a ring lattice pierced by an {\it effective magnetic field}. In particular, we shall see that our  system displays  signatures of the fractionalization phenomenon distinctive for  bright solitons.  
{\it Linear} chains of capacitivelly coupled qubits have been realized experimentally and have aroused considerable interest in the fields of driven-dissipative dynamics~\cite{ma2019A}, many-body localization~\cite{Roushan2017Spectroscopic,Xu2018emulating,guo2021observation,Chiaro2022direct}, disordered quantum phases of artificial matter ~\cite{mansikkamaki2021phases}, quantum walks~\cite{yan2019strongly, Giri2021Two,PainterQRWScience}, and perfect quantum state transfer~\cite{sun2018,roy2024}. 
In these systems, the low-energy modes are plasmons that can be excited by microwave photons and propagate through the capacitors. Such excitations behave as bosons located at the qubit's sites, and because of the non-linear inductance of the Josephson junctions in the qubits, they  can interact with each other. In contrast to the Cooper pair dynamics occurring in Josephson junctions arrays~\cite{fazio2001quantum}, here the boson-boson interaction is effectively {\it attractive} \cite{hacohen2015cooling,fedorov2021photon}.  The important observation, here, is that the natural regime in which  such systems work is the one of moderate  number of bosons $N_p$, thus  opening the door for analyzing the attractive Bose-Hubbard dynamics~\cite{naldesi2019rise,polo2020exact,polo2022quantum,blain2023soliton} in conditions that might be challenging  to reach in other physical platforms. 

Our capacitively coupled qubits ring system  is depicted in Fig.~\ref{Fig1}(a). In particular, we note that the effective magnetic  field cannot induce any actual matter-wave motion as it occur for example in cold atoms.  
Nevertheless, we  can implement  a synthetic magnetic field in our superconducting qubit system through a  Floquet driving~\cite{eckardt2005superfluid,lin2009synthetic,kolovsky2011creating,jotzu2014experimental,lim2017electrically} of the transmon frequencies.  The latter is suitably engineered to facilitate the operations to be carried out on the superconducting circuit system and  protect  the  transmons performances from decoherence.

\section{System and methods}

Transmons are weakly anharmonic quantum oscillators that are well described by attractive bosons~\cite{hacohen2015cooling,fedorov2021photon}. The latter consist of collective oscillations in the circuit and the attraction results from the weak anharmonicity provided by the Josephson potential in the regime $E_J\gg E_C$, with $E_J$ the Josephson energy and $E_C$ the charging energy.  In a ring configuration with $N_s$ capacitively-coupled transmons, as schematized in Fig.~\ref{Fig1}, the bosonic excitations can hop from one transmon to the other and realize a Bose-Hubbard model of attractive bosons in a lattice \cite{mansikkamaki2021phases,orell2019probing,orell2022collective}:
\begin{equation}
H_0=\sum_{j=1}^{N_s}\left[\omega n_j - \frac{U}{2} n_j(n_j-1)-J_0(a^\dag_{j+1}a_j+{\rm H.c.})\right],
\end{equation}
where $a_{N_s+1}\equiv a_{1}$, $n_j=a^\dag_ja_j$, $\omega\simeq\sqrt{8E_{C}E_{J}}-E_{C}$, $U\simeq E_{C}$, $J_0\simeq  (C_0/(C_J+C_g))\sqrt{E_CE_J/8}$, and we set $\hbar=1$ henceforth.  The charging energy of the transmons $E_C=e^2/2(C_J+C_g)$ is expressed in terms of the capacitance of the Josephson junction $C_J$ and the capacitance to ground of the superconducting islands $C_g$, with $C_0\ll C_J,C_g$ being the capacitance between nearest neighbor transmons. Assuming $(C_0/(C_J+C_g))\sqrt{E_J/(8E_C)}\ll 1$, we have $J_0\ll U\ll \omega$ (see \footnote{Supplementary Materials for details on the circuit, on the driving protocol, and on the driven-dissipative dynamics.} for details on the circuit Hamiltonian), implying strongly attractive interaction. We note that the transmon frequency can be modulated $\omega\to\omega(t)$ by  employing the so-called split-junctions, composed by a parallel of Josephson junctions threaded by a time-dependent external flux $\varphi_x(t)$. Special working points are the so-called 'sweet spots', that  correspond to maxima or minima of the energy versus $\varphi_x$. In this regimes the qubit properties are particularly stable against flux noise \cite{chirolli2006decoherence,krantz2019,kjaergaard2020superconducting,siddiqi2021}. 

Synthetic gauge fields  can be realized experimentally through Floquet modulation protocols~\cite{eckardt2005superfluid,lin2009synthetic,kolovsky2011creating,jotzu2014experimental}. The modulation is described by time-dependent frequencies $\omega_j(t)=\omega-\delta\omega_j(t)$, where $\delta\omega$ is periodic with period $T$ and we assume $2\pi/T \gg J_0,\, U$. Therefore, the Hamiltonian acquires an additional term
\begin{equation}
H(t)=H_0-\sum_{j}\delta\omega_j(t) n_j.
\end{equation}
We can perform the unitary transformation 
$U(t)=\exp\left[i\sum_jn_j\int^tdt'\delta\omega_j(t')\right]$ so that the effective Floquet Hamiltonian in the fundamental band is given by the time-averaged Hamiltonian $H_{\rm eff}=\left\langle U^\dag(t)H(t)U(t)-iU^\dag(t)\dot{U}(t)\right\rangle_T$, with $\langle\ldots\rangle_T\equiv \int_0^T\frac{dt}{T}\ldots$. The goal of the protocol is that $H_{\rm eff}$ acquires a  complex hopping amplitude $\displaystyle{J_0\to J_0 e^{i \frac{\Phi}{N_s}}}$,  accounting  for the Peierls substitution employed in  one-dimensional matterwave circuits in presence of a gauge field \cite{peierls1997theory}.  A finite Peierls phase has been achieved in two dimensional arrays of transmons in the hard-core boson limit  ($U\to\infty$) with a combination of static gradients and local sinusoidal modulations of $\omega$ that match the gradient steps \cite{Roushan2017Spectroscopic,rosen2024implementing,rosen2024flatband}. We notice that such protocol typically lifts the transmons away from their sweet spot  and that in the absence of static gradients, a sinusoidal driving has been demonstrated to yield  generic values of Peierls phases in $\{0,2 \pi\}$ at the price of suitably coloring the modulation with more than one frequency~\cite{dunlap1986dynamic,grossmann1992localization,holthaus1992collapse,eckardt2005superfluid,struck2012tunable,flach2000directed,periodically2007periodically,verdeny2014optimal,borneman2010application}. We point out that synthetic gauge fields have been also realized in systems of transmons by suitable sets of two-qubit gates in the hard-core boson limit     \cite{neill2021accurately,morvan2022formation}. However, such investigations do not access the bright soliton physics determined by interactions.

In the next section, we propose a specific Floquet protocol that leads to generic Peierls phases  in $\{0,2 \pi\}$ and  that can be also employed with the transmons at the sweet spot.  Our Floquet scheme employs the so called Leviton dynamics.

\section{Synthetic gauge field by Leviton protocol} 

\begin{figure}[t]
	\centering
    \includegraphics[width=1.0\linewidth]{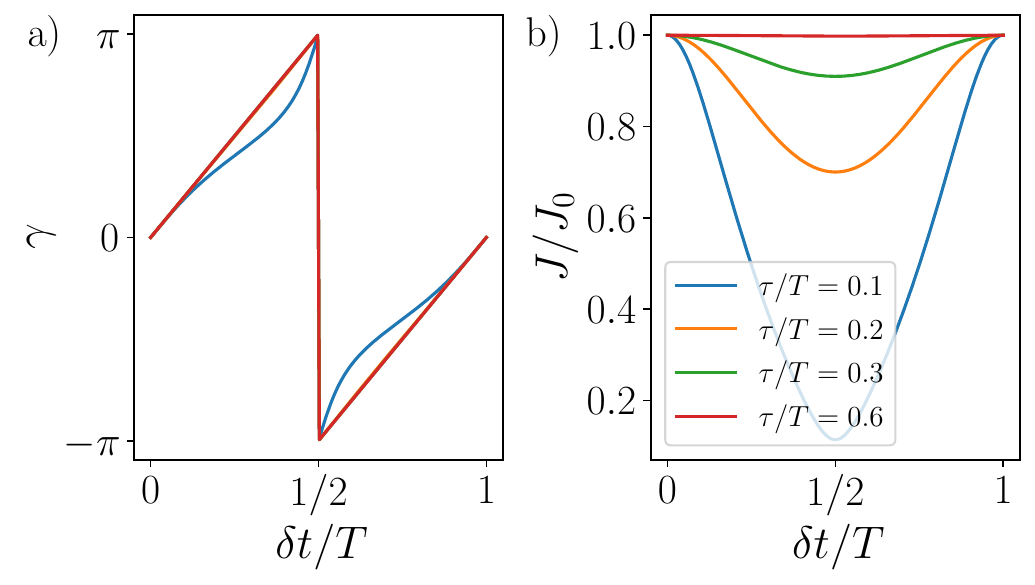}
	\caption{(a) Phase and (b) modulus of the complex hopping between two transmons modulated through Levitons, as a function of the time delay $\delta t/T$ for four different values of the Lorentzian width $\tau/T$ [in (a), only the curve for the smallest width can be distinguished from the other three curves].}
    \label{Fig2}     
\end{figure}

Levitons consist of suitable Lorentzian pulses that are characterized by an exponential Fourier spectrum and yield a quantized phase advanced of $2\pi$. Based on such a feature, Levitons have been originally introduced in quantum physics   by Levitov as noise-free electronic excitations for edge states in the Quantum Hall effect regimes \cite{levitov1996electron,ivanov1997coherent,keeling2006minimal,dubois2013minimal,glattli2017levitons}.  Here, we propose Levitons to implement a synthetic gauge field by Floquet dynamics. The protocol consists in modulating  a suitable subset of adjacent $N_m<N_s$ transmons  through a sequence of Leviton pulses of period $T$ and width $\tau$
\begin{equation}
    \label{Eq:domega}\delta\omega_j(t)=\sum_k\frac{2\tau}{(t-t_j-kT)^2+\tau^2},
\end{equation}
in which $t_j$ is a site-dependent reference time. We shall see that $N_m\geq 2$ is a necessary condition for the protocol to work. Thanks to the properties of Levitons, it can be analytically shown that the hopping between the modulated adjacent transmons acquires a phase $\gamma$
\begin{equation}
Je^{-i\gamma}=\langle J_0e^{-i\int^tdt' (\delta\omega(t')-\delta\omega(t'-\delta t))}\rangle_T,
\end{equation}
that can be adjusted by varying the time delay $\delta t=(t_{j+1}-t_j)$ between pulses in adjacent transmons. This feature can be explicitly seen for $e^{-4\pi\tau/T}\ll 1$ for which we have $J e^{-i\gamma}\simeq J_0 e^{-2\pi i\delta t/T}$ (see also Supplemental material). Indeed, Fig.~\ref{Fig2}(a) shows an approximately linearly increasing phase $\gamma$ between two adjacent transmons versus $\delta t$, for different values of the normalized Lorentzian width $\tau/T$ such that $e^{-4\pi\tau/T} \ll 1$. At the same time,  the hopping rate $J/J_0$ is also shown to be weakly dependent on $\delta t$, as shown in  Fig.~\ref{Fig2}(b).

We comment that, due to the linear dependence of the phase $\gamma$ on  $\delta t$, modulating all the transmons with time-shifted pulses does not lead to any synthetic flux (since all phases gained cancel out around the ring). In turn, by modulating only a subset of $N_m$ transmons  a net synthetic flux can be achieved.  This choice introduces two ``external'' links between the modulated and non-modulated transmons, and the associated hopping rate $J'$  is suppressed:  
\begin{equation}
    J'=\langle J_0 e^{-i\int^tdt' \delta\omega(t'-t_j)}\rangle_T= -J_0e^{-2\pi\tau/T}.
\end{equation}
Therefore, in order to preserve the translation invariance of the system in the fundamental Floquet band, the bare hopping at the external links needs to be modified to compensate the suppression due to the modulation, $J_0\to J_0'=J_0 e^{2\pi\tau/T}$. Additionally, the $\pi$ phase shift acquired in one external link is  compensated by an analogous one at the second external link.  The result of the protocol is that we can impart a net synthetic flux to the system
\begin{equation}\label{Eq:gamma-vs-dt}
\Phi=2\pi(N_m-1)\delta t/T,  
\end{equation} 
which can be controlled with the time delay of the pulses between adjacent transmons.

Several comments are now in order. (1) The Floquet protocol based on Leviton dynamics can interestingly be employed at the transmon sweet spots. To see this, we notice that the minimum of the modulation $\delta\omega(t)$ is $\omega_c=(2\pi/T) \tanh(\pi \tau/T)$ and when the transmons are  modulated at the sweet spot they naturally generate a 'dc component' $\omega_c$ \cite{Note1}. (2) In order to utilize the series of Levitons at the sweet spots, it is then necessary to shift the frequency $\omega$ of the modulated transmons by $\omega_c$, so to  compensates the dc component. The proper value of the Floquet period $T$ and the width of the Lorentzians $\tau$ can then be adjusted a posteriori, so to match the step $\omega_c$ and the bare hopping $J_0'$. (3) We note that the rescaling of $J_0'$ breaks the original translation invariance of the Hamiltonian $H_0$ and can introduce coupling to higher Floquet bands in the exact time-dependent dynamics. Nevertheless, for sufficiently high Floquet frequency such that $2\pi/T \gg J_0'e^{2\pi \tau/T}$ these transitions can be suppressed \cite{Note1}. 
(4) Finally, we point out that requiring the total amplitude of the modulation to be a fraction of the transmon frequency sets the correct hierarchy of frequency scales: $2\pi/T<2\pi/\tau<\omega$. This is not in contrast with the requirement that the Floquet frequency is effectively the higher scale in the system as in each sector of constant number of bosons the dynamics is controlled by $J/U$; typical values of $J/\omega\sim 10^{-2}$ can fully justify a Floquet  regime in which $J\ll 2\pi/T\ll \omega$. The numerical comparison between the target and the actual dynamics of the system is carried out in the Supplemental material \cite{Note1}.

\section{Spectrum measurement}

Having established a protocol for imparting  a synthetic flux to the system, we now propose an experimentally  feasible scheme  for detecting the effect of $\Phi$ on the superconducting circuit. Specifically, we consider  a transmission line capacitively coupled to the ring of transmons and add the driving term $\Omega\cos(\omega_dt)(a_1+a_1^\dag)$ to the Hamiltonian, where $\Omega$ characterizes the strength of the driving.  The ring network can then be coherently  driven by microwaves of frequency $\omega_d$ and by the same transmission line  the outgoing photons can be monitored. This way, we can map-out the spectrum  of the system through the amplitude of the reflected wave $S(\omega_d)$. Following Ref.~\cite{fedorov2021photon} and by time averaging the combined Floquet modulation and external driving dynamics (in the rotating frame of the driving), the effective Hamiltonian becomes \cite{Note1}
\begin{eqnarray}\label{Eq:Heff}
H_{\rm eff}&=& \sum_{j=1}^{N_s}\left[(\omega-\omega_d) n_j - \frac{U}{2} n_j(n_j-1)\right]\nonumber\\
&-&J_0\sum_{j=1}^{N_s}(e^{i\Phi/N_s}a^\dag_{j+1}a_j+{\rm H.c.})+\frac{\Omega}{2}(a_1+a_1^\dag),
\end{eqnarray} 
and we have neglected fast rotating terms for $\Omega\ll\omega,\,\omega_d$. Clearly, the coherent driving implies  non-number-conserving process (it couples sectors with different number of bosons): The system coherently absorbs photons from the transmission line and incoherently loses photons both via the transmission line and by relaxation processes in each transmon; in addition, dephasing in each transmon can also result in a loss of coherence. We expect that solitonic bands with different $N_p$ are separated in energy by $\sim U$. Therefore, in order to avoid band mixing, we require $\Omega\ll U$. For simplicity, we assume that only transmons that are not modulated through the Floquet protocol are connected to the external drive.

We study the evolution of the entire system phenomenologically through a Lindblad master equation~\cite{breuer-petruccione2002,biella2015photon,fazio2024,venkatraman2024}
\begin{equation}
\frac{\partial \hat{\rho}}{\partial t}={\cal L}[\hat\rho]=-i[H_{\rm eff},\hat{\rho}]+{\cal D}_1[\hat{\rho}]+{\cal D}_\phi[\hat{\rho}],
\end{equation}
with  ${\cal D}_\alpha[\hat\rho]=\sum_{j}\left[O_{j,(\alpha)}\hat{\rho}O_{j,(\alpha)}^\dag-\frac{1}{2}\{O^\dag_{j,(\alpha)}O_{j,(\alpha)},\hat{\rho}\}\right]$ a general dissipator, which for relaxation processes is specified by $O_{j,(1)}=\sqrt{\gamma_1}a_j$ and for dephasing processes is described by $O_{j,(\phi)}=\sqrt{\gamma_\phi}a^\dag_ja_j$, with $\gamma_1$ and $\gamma_\phi$ the transmon relaxation and pure dephasing rates, respectively, which within a phenomenological approach should be interpreted as effective ones. To model the system output, we follow Refs.~\cite{hacohen2015cooling,biella2015photon,fedorov2021photon} and assume that the site $1$ couples with rate $\Gamma$ to the transmission line. Following a quantum Langevin approach \cite{collett1984squeezing,breuer-petruccione2002,mirhosseini2019cavity}, coherent irradiation results in the input field mode amplitude being related to the driving strength $\Omega$ as $\sqrt{\Gamma}\langle a^\dag_{\rm in}\rangle\approx i\Omega$ and the output field is $a^\dag_{\rm out}\approx \sqrt{\Gamma}a^\dag_1$. In this regime, photon reflection amplitude is then expressed as 
\begin{equation}
    S(\omega_d)=\langle a^\dag_{\rm out}\rangle/\langle a^\dag_{\rm in}\rangle=\Gamma {\rm Tr}[\hat{\rho}_{ss}a^\dag_1]/i\Omega\, 
\end{equation}
where $\hat{\rho}_{ss}$ is the steady-state density matrix satisfying $\partial \hat{\rho}_{ss}/\partial t=0$. We then perform numerical simulations for system of size $N_s=4$, keeping up to $N_p=3$ excitations, and we choose experimentally relevant parameters taken from Ref.~\cite{fedorov2021photon}: $\omega=3.9~{\rm GHz}$, $U=0.188~{\rm GHz}$, $J=0.041~{\rm GHz}$.

\begin{figure}[t]
	\centering
    \includegraphics[width=1.0\linewidth]{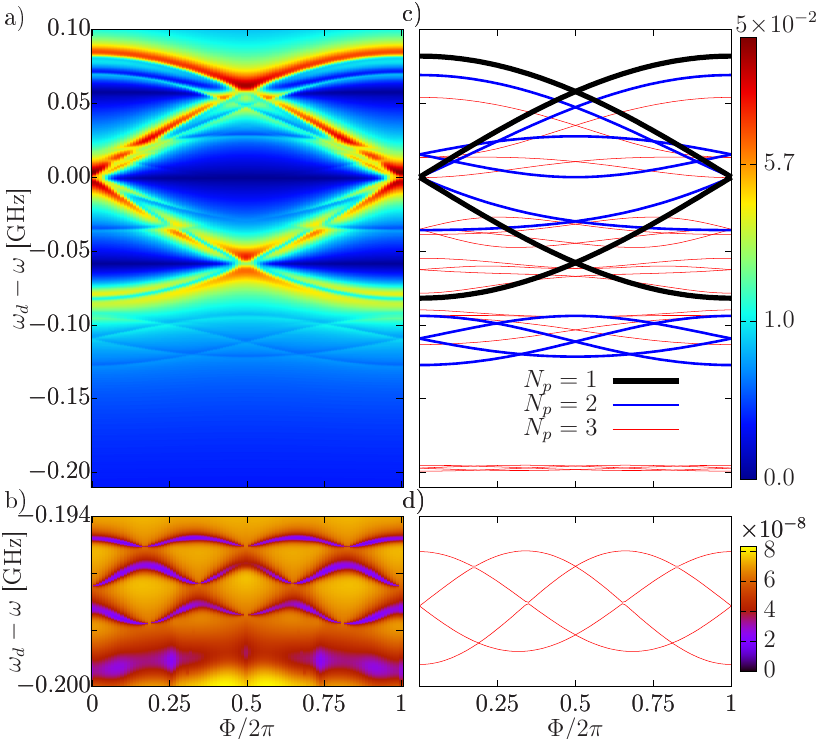}
	\caption{(a) Modulus of the wave reflected off the driven system as a function of the synthetic flux $\Phi$, detuning $\omega_d-\omega$ (in GHz), and driving amplitude $\Omega = 0.01$~GHz, for a chain of $N_s=4$ transmons. A clear single-particle spectrum ($N_p=1$) appears as a broad bright signal around zero detuning and shows the expected $2\pi$ periodicity as a function of the synthetic flux. At negative detuning, a band made of two-particle bound states characterizing bright solitons appears with the predicted periodicity halved to  $\pi$ (lower part of the spectra). (b) For further increasing the external drive $\Omega = 0.04$~GHz three-particle bound states emerge with one third periodicity. (c) Superimposed spectra of the $N_p=1,2,3$ sectors, showing the correspondence with the output signal. (d) Zoom of the $N_p=3$ sector. The rest of the system parameters are given in the main text. The color bar in a) is made non-linear by setting $z(x,y)^{2/5}$.}
    \label{Fig3}    
\end{figure}

In Fig.~\ref{Fig3}(a) we present the results for weak driving $\Omega=0.01~{\rm GHz}$, $\Gamma=5~{\rm MHz}$, $\gamma_1=1~{\rm MHz}$, and we neglect $\gamma_\phi$ (see \cite{Note1} for a discussion of the role of $\gamma_\phi$). Notice that typical decoherence rates are dominated by relaxation processes and that the chosen value of $\gamma_1$ is a worst-case scenario rather than an optimistic value. A clear broad bright signal following the dispersion $E^{(1)}_k=\omega-\omega_d-J\cos(2\pi k/N_s+\Phi/N_s)$, for $k=0,\ldots, N_s-1$ emerges. This feature, displaying the full $2\pi$ flux periodicity, is identified as corresponding to the $N_p=1$ sector. Remarkably, a weaker signal in the form of a dip is visible, in lower part of Fig.~\ref{Fig3}(a), which corresponds to the $N_p=2$ sector. Here, we see the inception of the reduced periodicity, where a minimum appears between $\Phi=0$ and $\Phi=2\pi$. The apparent deviations with respect to the perfect reduced periodicity were demonstrated to arise as $1/N_s$ corrections \cite{piroli2016local,naldesi2019rise,naldesi2022enhancing}. In order to observe the $N_p=3$ bound state  we needed to increase the driving amplitude $\Omega$ to be of order $J$. We comment that  the time scale for photon hopping from the transmission line to the lattice site $j=1$, determined by $2\pi/\Omega$, has to be comparable to or shorter than the time scale on which photons hop from one transmon to the other, $1/J$, requiring $\Omega\gtrsim J$. For $\Omega=0.04~{\rm GHz}$, $\Gamma=0.1~{\rm MHz}$, and $\gamma_1=10^{-2}~{\rm MHz}$ the sector $N_p=3$ appears, as shown in Fig.~\ref{Fig3}(b). The lowest energy band is displayed as significantly broadened and we can only observe the fractional $2\pi/3$ periodicity in the excited bands. We also point out that degeneracies in the level crossings of the many-body spectrum, taking place at specific values of the synthetic flux, are lifted as a combined effect of driving and dissipation. 

Our numerical findings are corroborated by a perturbative calculation  valid for large $U$ of the  energy dispersion~\cite{Note1}
\begin{equation}
    E^{(N_p)}_k=E_0^{(N_p)}-J_{(N_p)}\cos\left(2\pi k/N_s+N_p\Phi/N_s\right),
\end{equation}
 with $E_0^{(N_p)}=N_p(\omega-\omega_d)-\frac{U}{2}N_p(N_p-1)$ and $J_{(N_p)}=\frac{J N_p}{(N_p-1)!}(J/U)^{N_p-1}$ \cite{mansikkamaki2022beyond}. This formula is in good agreement with the exact spectrum shown in Figs.~\ref{Fig3}(c) and (d).

In summary, our analysis  shows that {\it the low energy bands in the absorption spectrum have a periodicity reduced by $N_p$ reflecting the size of the bound state  formed in the system.} According to the predictions in Ref.~\cite{naldesi2019rise}, for sufficiently strong attractive interactions the energy bands corresponding to bound states detach from the single particle (scattering) band [see Fig.~\ref{Fig3}c)]. We note that solitons are increasingly more fragile with $N_p$ to decoherence. This is due to the fact that the solitons take approximately a time $\sim N_s/J_{(N_p)}$ to go around the ring of transmons. Therefore, to observe coherent flux oscillations such a time has to be shorter than the single transmon relaxation time $\sim 1/(N_p\gamma_1)$ and pure dephasing time $1/\gamma_\phi$ \cite{Peterer2015}. Thus we require $N_p\gamma_1,\,\gamma_\phi\ll J_{(N_p)}/N_s$, so that the protocol needs a ratio $J/U$ not too small. The chosen value of the  relaxation rate in Fig.~\ref{Fig3}(b) is within the state-of-the-art values \cite{milul2023,ganjam2024} and we have checked that the solitons and their periodicity are visible up to $\gamma_1=0.1~{\rm MHz}$ (see \cite{Note1} for an assessment of the coherence properties).

\section{Conclusions} 

We have shown how a ring-shaped system of  capacitively-coupled transmons can be employed to study the many-body dynamics of attracting bosons  in presence of an effective magnetic field. The latter is obtained through a suitable Floquet driving  protocol employing Lorentzian pulses (Levitons). This way,  we widen the scope of the  superconducting circuits quantum simulators significantly. 

The bosonic effective dynamics is indeed  characterized by the formation of bright solitons whose smoking gun    
is  the  fractionalization of flux quantum that is predicted to be displayed in the `band structure' of the system (energy versus magnetic field)~\cite{naldesi2019rise,naldesi2022enhancing}. Such flux quantum fractionalization emerges  for realistic experimental conditions ~\cite{fedorov2021photon}, in a suitable  reflection protocol (see Fig.~\ref{Fig3}). 

Important problems such as the impact of disorder on the fractionalization, macroscopic phase coherence and SQUID physics \cite{polo2022quantum}, Aharonov-Bohm oscillations in bosonic systems \cite{haug2019aharonov}, $N$-component bosons, and mesoscopic simulation of lattice gauge theories~\cite{domanti2024aharonov}, can be studied in a regime of parameters that is complementary to the working conditions of strongly interacting cold atoms systems.  Such a platform is also complementary to others describing synthetic gauge fields in weakly-coupled photons, such as those occurring in topological photonics \cite{ozawa2019}.
Relying on the relevance of solitons in quantum metrology \cite{cornishquantum2009,mcdonald2014bright,helmsagnac2015,marchantquantum2016,naldesi2023massive},  we expect our scheme can open new perspectives for superconducting circuits-based  quantum sensing.

\section{Acknowledgements} 
We thank B. Blain, E. C. Domanti, and G. Marchegiani for discussions.

\bibliography{BH-bib}

\clearpage
\onecolumngrid

\begingroup
\leftskip=0cm plus 0.5fil
\rightskip=0cm plus -0.5fil
\parfillskip=0cm plus 1fil
    \textbf{\large Supplemental Material for:} \\
    \textbf{\large ``Synthetic fractional flux quanta in a ring of superconducting  qubits''}
   \par
\endgroup
\vspace{0.5cm}

\setcounter{section}{0}
\setcounter{equation}{0}
\setcounter{figure}{0}
\setcounter{table}{0}
\setcounter{page}{1}
\makeatletter
\renewcommand{\theequation}{S\arabic{equation}}
\renewcommand{\thefigure}{S\arabic{figure}}
\renewcommand{\thesection}{S\Roman{section}}

\twocolumngrid

\section{The system Hamiltonian}

We consider a chain of $N_s$ capacitively coupled transmons. Each node of the chain is composed by a superconducting island with phase $\phi_j$ and it is coupled to the ground via a transmon qubit, that is composed by a Josephson junction with energy $E_{J,j}$, capacitively shunted by a capacitance $C_{J,j}$. The superconducting island is capacitively coupled to the other transmons at the left and the right along the chain via capacitances $C_{0,j-1}$ and $C_{0,j}$, respectively, and on the top to a voltage source, that is typically realized via a transmission line resonator, via capacitance $C_{g,j}$. The chain is closed via an additional capacitance $C_{0,N}$.

The Lagrangian ${\cal L}$ of the circuit is a function of the $\phi_j$ and their time derivative $\dot{\phi}_j$, ${\cal L}(\{\phi_j\},\{\dot{\phi}_j\})$, and it is given by the difference between the charging energy $K$ and the Josephson potential $U$, 
\begin{equation}\label{Eq:FullU}
{\cal L}=K(\{\dot{\phi}_j\})-U\left(\{\phi_j\}\right).
\end{equation}
where the charging energy $K$ is 
\begin{eqnarray}
K&=&\frac{1}{2}\sum_{j=1}^{N_s}\left[C_{J,j}\dot{\Phi}_n^2+C_{g,j}(\dot{\Phi}_j-V_j)^2\right]
\nonumber\\
&+&\frac{1}{2}\sum_{j=1}^{N_s}C_{0,j}(\dot{\Phi}_j-\dot{\Phi}_{j+1})^2
\end{eqnarray}
where $\Phi_j=\Phi_0\phi_j/2\pi$, $\Phi_0=h/2e$, and $\phi_{N+1}\equiv \phi_1$. 
$K$ can be compactly written in terms of a capacitance matrix, whose entries are ${\cal C}_{j,j}=C_{J,j}+C_{g,j}+C_{0,j-1}+C_{0,j}$, ${\cal C}_{j,j+1}={\cal C}_{j+1,j}=-C_{0,j}$, for $n=1,\ldots,N$, and ${\cal C}_{N,1}={\cal C}_{1,N}=-C_{0,N}$, so that it takes the compact form
\begin{equation}
K=\frac{1}{2}\sum_{i,j=1}^{N_s}\dot{\Phi}_j{\cal C}_{i,j}\dot{\Phi}_j-\sum_{j=1}^{N_s}C_{g,j}\dot{\Phi}_jV_j.
\end{equation}The Josephson potential is 
\begin{equation}\label{Eq:FullK}
U=\sum_{j=1}^{N_s}E_{J,j}(1-\cos(\phi_j)).
\end{equation}
The Hamiltonian is obtained by performing a Legendre transform and, through definition of the charges $q_i=\partial {\cal L}/\partial \dot\Phi_i=\sum_j{\cal C}_{i,j}\dot{\Phi}_j-C_{g,i}V_i$, it reads
\begin{equation}
H=\frac{1}{2}\sum_{i,j=0}^{N_s}(q_i+q_{g,i})[{\cal C}^{-1}]_{i,j}(q_j+q_{g,j})+U(\{\phi_j\}),
\end{equation}
where $q_{g,i}=C_{g,i}V_i$ are offset charges on the superconducting islands. The latter can also include time-dependent voltages $q_{g,n}(t)=\sum_m[{\cal C}^{-1}]_{n,m}C_{g,m}V_m(t)$ that account both for the electric field of the transmission line resonator and fluctuating charges that arise due to dielectric losses.

The capacitance matrix can be analytically inverted in the case of a uniform closed chain. By going to Fourier space we have $C(k)=2C_0(\eta+1-\cos(k))$, with $\eta=(C_J+C_g)/C_0$ and $k=2\pi n_k/N_s$, so that the matrix elements of the inverse capacitance matrix read
\begin{equation}
{\cal C}^{-1}_{i,j}=\frac{z_-^{|i-j|}}{C_0(z_+-z_-)}=\frac{e^{-|i-j|/\ell_s}}{2C_0\sqrt{2\eta+\eta^2}},
\end{equation}
with $z_\pm=1+\eta\pm \sqrt{(1+\eta)^2-1}$. Being $z_-\leq 1$ the interaction falls off exponentially with a screening length $\ell_s=-1/\ln(z_-)$. We are interested in short range capacitive coupling, so we require  $\eta\gg 1$ and truncate the charge-charge interaction to nearest neighbor, so to have a collection of transmons coupled by nearest neighbor capacitive coupling. 

Each transmon is described by the Hamiltonian 
\begin{equation}
H_{j}=4E_{C,j}N^2_j-E_{J,j}\cos(\phi_j),
\end{equation}
where $N_j=q_j/2e=-i\partial/\partial\phi_j$ is the number operator conjugate to the phase, $[\phi_j,N_j]=i$. The transmons have in principle all slightly different Josephson energy $E_{J,j}$ and charging energy $E_{C,j}=e^2/(2(C_{J,j}+C_{g,j}))$. Nevertheless, each Josephson junction can be realized as a parallel of two nominally equal junctions (SQUID), that can be tuned through external fluxes both in a static and time-dependent way. We then assume the transmon to be nominally equal and introduce bosonic operators describing plasmonic modes $\phi_j=\frac{\ell}{\sqrt{2}}(a_j+a^\dag_j)$ and $N_j=\frac{-i}{\ell\sqrt{2}}(a_j-a^\dag_j)$, with $\ell=(8E_{C}/E_{J})^{1/4}$. By expanding the Josephson potential up to fourth order in the phases and retaining only number-conserving terms, we obtain the Hamiltonian
\begin{equation}
H_0=\sum_{j=1}^{N_s}\left[\hbar \omega n_j -\frac{U}{2}n_j(n_j-1)\right]+J_0\sum_{n=1}^{N_s}(a^\dag_{j+1}a_j+{\rm H.c.}),
\end{equation}
with $a_{N_s+1}=a_1$,  $n_j=a^\dag_ja_j$, $\hbar\omega=\sqrt{8E_{C}E_{J}}-E_{C}$, $U=E_C$, and 
\begin{equation}
    J_0=\frac{2E_C  z_-}{\ell^2\sqrt{1+2/\eta}}\simeq \frac{E_C}{\eta\ell^2}=\frac{C_0}{C_J+C_g}\sqrt{E_JE_C/8},
\end{equation}
so that for $C_0/(C_J+C_g)\sim 10^{-2}$ we obtain a hopping term that is two orders of magnitude smaller than the transmon frequency $\omega$.

\subsection{Time-dependent modulation}

We assume to have split-transmons, in which the Josephson junction is replaced by a SQUID, the parallel of two junctions with energy $E_{J1}$ and $E_{J2}$ that enclose a normalized flux $\varphi_{x}(t)=2\pi\Phi_x(t)/\Phi_0$, with $\Phi_0=h/2e$ the superconducting flux quantum and $\Phi_x(t)$ the external flux, which we assume to be time-dependently driven. The effective junction potential energy can be written as
\begin{equation}
U_{J}=-E_{J}(t)\cos(\phi-\theta_{x}(t)),
\end{equation}
where
\begin{eqnarray}
E_{J}(t)&=&\sqrt{E_{J1}^2+E_{J2}^2+2E_{J1}E_{J2}\cos(\varphi_x(t))}\nonumber\\
\tan(\theta_x(t))&=&\frac{E_{J1}-E_{J2}}{E_{J1}+E_{J2}}\tan\left(\frac{\varphi_x(t)}{2}\right).
\end{eqnarray}
The phase $\theta_x(t)$ can be removed from the Josephson energy by performing a unitary transformation 
\begin{equation}
U(t)=e^{iN\theta_x(t)}=D(\theta_x/(\ell \sqrt{2}))
\end{equation}
where $D(\alpha)=e^{\alpha a^\dag-\alpha^*a}$ is a bosonic displacement operator satisfying $D^\dag(\alpha)=D(-\alpha)$ and $D^\dag(\alpha)a D(\alpha)=a+\alpha$. The new Hamiltonian reads
\begin{equation}
H'(t)=U^\dag(t)H(t)U(t)-iU^\dag(t)\dot{U}(t)
\end{equation}
and we have
\begin{equation}
H'(t)=4E_CN^2-E_J(t)\cos(\phi)+\dot{\theta}_x N.
\end{equation}
For practical purposes we choose to have a symmetric SQUID and assume $E_{J2}=E_{J1}\equiv E_J/2$, so that $\theta_x=0$. In addition, we assume the amplitude of the flux modulation to be sufficiently small. If we expand around $\varphi_x=0$ we can write
\begin{equation}
E_J (\varphi_x)=E_{J}\left(1-\frac{\varphi^2_x}{8}\right).
\end{equation}

We then quantize the time-independent part of the Hamiltonian, by defining $\ell=(8E_C/E_{J})^{1/4}$, $\hbar\omega=\sqrt{8E_CE_{J}}-E_C$ and write the Hamiltonian as
\begin{equation}
H_T = \omega n-\frac{E_C}{2}n(n-1)-\frac{\delta\omega(t)}{2}(a+a^\dag)^2,
\end{equation}
with $\delta\omega(t)=E_J\varphi_x^2(t)\ell^2/16 =\omega\varphi_x^2(t)/16$. By neglecting non-number-conserving terms we can write
\begin{equation}
H_T = \left[\omega-\delta\omega(t)\right]n-\frac{E_C}{2}n(n-1).
\end{equation}
This analysis is valid for $\varphi_x\ll 1$, implying that the modulation $\delta\omega$ is considered as a small fraction of the unperturbed frequency $\omega$. 

At the same time, as it will be clear in the next section, the ideal pulse is constituted by a series of Lorentzian peaks, whose amplitude is in general not too small. We then consider higher order in both $\varphi$ and $\varphi_x$ and write
\begin{eqnarray}
    H_T&=&\omega n-\frac{E_C}{2}n(n-1)-\frac{\delta\omega(t)}{2}(a+a^\dag)^2\nonumber\\
    &+&\frac{\delta\omega(t)\ell^2}{48}(a+a^\dag)^4,
\end{eqnarray}
with $\delta\omega(t)=\omega(\varphi_x^2(t)-\varphi_x^4(t)/48)/16$, and retaining only number-conserving terms we obtain
\begin{equation}
    H_T=\left[\omega-\delta\omega(t)(1-\ell^2/4)\right]n-\frac{E_C}{2}\left(1-\frac{\ell^2\delta\omega(t)}{4E_C}\right)n(n-1).
\end{equation}
We see that the main picture holds with the additional contribution of a time-dependent modulation of the interaction term. The latter can be averaged over time in the Floquet scheme and amounts to an effective decrease of the absolute value of the interaction strength by a factor of order $(1-2\delta\omega/\omega)$.

\section{Leviton pulses}

In the main text we presented results for the synthetic gauge field obtained through Leviton pulses. Here, we present details of the calculations and  provide the full modulation and driving protocol. 

We assume to apply the modulation to a subset of adjacent transmons of the chain such that the Hamiltonian reads
\begin{equation}
H(t)=\tilde{H}_0-{\sum_{j}}'\delta\omega_j(t)a^\dag_j a_j.
\end{equation}
where the prime signifies that the sum is restricted to a
subset, and the tilde that in the bare Hamiltonian $H_0$ we modify the hopping amplitude at the external link as $J_0\to J_0'$, so to obtain an effective Hamiltonian in the fundamental Floquet band that is translational invariant, as explained in the main text. 

We modulate the selected transmons with a train of Lorentzians, such that 
\begin{eqnarray}
    \delta\omega(t)&=&\sum_k\frac{2\tau}{(t-kT)^2+\tau^2}\\
    &=&\frac{2\pi}{T}+\frac{4\pi}{T}\sum_{n=1}^\infty e^{-2\pi n\tau/T}\cos(2\pi n t/T)\\
    &=&\frac{i\pi}{T}(\cot(\pi(t+i\tau)/T)-\cot(\pi(t-i\tau)/T)),~~~~~~
\end{eqnarray}
and  $\delta\omega_j(t)=\delta\omega(t-t_j)$ for some reference time $t_j$.  We then define the dynamical phase
\begin{equation}
f_j(t)= \int^tdt'\delta\omega_j(t')\equiv f(t-t_j),
\end{equation}
where 
\begin{equation}
f(t)=i\left[\ln(-\sin(\pi (t+i\tau)/T))-\ln(\sin(\pi (t-i\tau)/T))\right],
\end{equation}
and the unitary transformation yields a dynamical phase for the bosons,
\begin{equation}
a_j\to a_j e^{if_j(t)}.
\end{equation}
Focusing on the case of only two adjacent transmons modulated, say $j$ and $j+1$, the relevant complex hoppings in the fundamental Floquet band acquire the form
\begin{eqnarray}
    J_{j,j+1} e^{i\gamma_{j,j+1}}&=& J_0{\cal I}_1(\delta t),\\
    J_{j-1,j}e^{i\gamma_{j-1,j}}&=&J_0'{\cal I}_2,\\
    J_{j+1,j+2}e^{i\gamma_{j+1,j+2}}&=& J_0'{\cal I}_2^*.
\end{eqnarray}
where $\delta t=t_{j+1}-t_{j}$ and 
\begin{eqnarray}
{\cal I}_1(\delta t)&=&\int_0^{T}\frac{dt}{T}e^{-if(t)+if(t-\delta t)}\nonumber\\
{\cal I}_2&=&\int_0^{T}\frac{dt}{T}e^{-if(t)},
\end{eqnarray}
It follows that the total synthetic flux is
\begin{equation}
\Phi={\rm Arg}[{\cal I}_1(\delta t)],
\end{equation}
and the renormalized hopping rates are
\begin{eqnarray}
J_{j,j+1}&=&J_0\left|{\cal I}_1(\delta t)\right|,\\
J_{j-1,j}&=& J_{j+1,j+2} = J_0'\left|{\cal I}_2\right|.
\end{eqnarray}
The integrals ${\cal I}_1$ and ${\cal I}_2$ can be done analytically in the complex plane,
\begin{eqnarray}
    {\cal I}_1&=&1-\frac{2\sinh(2\pi\tau/T)\sin(\pi\delta t/T)}{\sin(\pi(\delta t-2i\tau)/T)},\label{Eq:I1}\\
    {\cal I}_2&=&-e^{-2\pi\tau/T}\label{Eq:I2}.
\end{eqnarray}
An expansion for $e^{-4\pi\tau/T}\ll 1$ gives
\begin{equation}
    {\cal I}_1\simeq e^{-2\pi i\delta t/T}\left[1-4e^{-4\pi\tau/T}\sin^2(\pi\delta t/T)\right].
\end{equation}
By inspection of the expressions for the modulation $\delta\omega(t)$ we see that its amplitude $\delta\tilde{\omega}$ is given by
\begin{equation}
    \delta\tilde{\omega} =\frac{2}{\tau}\frac{2\pi\tau/T}{\sinh(2\pi\tau/T)}.
\end{equation}
In order for $\delta\omega(t)$ to be a small perturbation for the modulated transmons, we need $\delta\tilde{\omega}/\omega\ll 1$. Moreover, we want the renormalization of the hoppings to be relatively small and the Floquet modulation to be fast on the scale of the hopping. It follows that the correct hierarchies of scales are
\begin{equation}\label{eq:scales}
\frac{2\pi}{T}<\frac{2\pi}{\tau}<\omega,\qquad J_0\ll \frac{2\pi}{T}\ll\omega.
\end{equation}

\subsection{Coupling to Floquet side bands}

In order to check the goodness of the Floquet approximation, it is important to estimate the coupling to additional side bands located at energies shifted by multiples of the Floquet frequency $\omega_f=2\pi/T$. Introducing $\kappa=e^{-2\pi\tau/T}$, the complex hopping are
\begin{eqnarray}
    J'_{(n)}/J_0'&=&\int_0^T\frac{dt}{T}e^{2\pi i n t/T}e^{-if(t)}\\
    &=&\int \frac{dz}{2\pi i}z^{n-1}\frac{1-\kappa z}{z-\kappa}\\
    &=&\begin{cases}
    \kappa^{n-1}-\kappa^{n+1}, & \text{if $n\geq 1$},\\
    -\kappa, & \text{if $n=0$},\\
    0 & \text{if $n<0$},
  \end{cases}
\end{eqnarray}
and
\begin{eqnarray}
    J_{(n)}/J_0&=&\int_0^T\frac{dt}{T}e^{2\pi i nt/T}e^{-if(t)+if(t-\delta t)}\\
    &=&\frac{1}{\kappa}\int_C\frac{dz}{2\pi i}z^{n-1}\frac{(1-\kappa z)(z-\kappa w)}{(z-\kappa)(w/\kappa -z)}\\
    &=&\begin{cases}
    \kappa^{n}\frac{(1-\kappa^2)(1-w)}{w-\kappa^2}, & \text{if $n\geq 1$},\\
    1+\frac{(1-\kappa^2)(1-w)}{w-\kappa^2}, & \text{if $n=0$},\nonumber\\
    \left(\frac{\kappa}{w}\right)^{-n}\frac{(1-\kappa^2)(1-w)}{w-\kappa^2}& \text{if $n<0$},
  \end{cases}
\end{eqnarray}
with $w=e^{2\pi i\delta t/T}$. The coupling to higher Floquet bands mediated by $J_{(n)}$ is always perturbatively small in $\kappa$, so that it can be safely neglected. In turn, the coupling to the first Floquet band for the rescaled case $J_0'=J_0/\kappa$ is of order $1/\kappa$, so that the first Floquet band is accessible to the dynamics and the corrections are on order $TJ_0^2/\kappa^2$. It follows that the first Floquet band can be neglected for $T\ll \kappa^2/J_0$, in general a more restrictive condition that the one given in Eq.~\eqref{eq:scales}.

\subsection{Working at the sweet spot of flux noise insensitivity.}

In order to minimize the dephasing processes, it is important to work with all transmons at their sweet spot for flux noise insensitivity. Indeed, random fluctuations of the flux $\varphi_x$ that controls the frequency $\omega$ yield dephasing \cite{chirolli2006decoherence,krantz2019,kjaergaard2020superconducting,siddiqi2021}.  From the dependence of the transmons' Josephson energy on the flux $\varphi_x$, we see that these points represent maxima or minima of $E_J(\varphi_x)$, in a way that their first derivative with respect to flux is zero, and the transmons are sensitive only at second order. Modulating the flux $\varphi_x(t)$ at the sweet spot then results in a time-dependent frequency $\delta\omega(t)$ that is a quadratic function of the flux, so that it has a finite dc component. Focusing on modulation around a maximum of $E_J$, the Leviton pulses are then described by
\begin{equation}
    \delta\omega(t)=\sum_k\frac{2\tau}{(t-kT)^2+\tau^2}-\omega_c,
\end{equation}
where $\omega_c=(2\pi/T)\tanh(\pi\tau/T)$. First of all, we point out that in the regime $\omega T\gg 2\pi$ and $\tau/T<1$ but not too small, the frequency shift is much smaller than the transmon frequency,  $\omega_c/\omega\ll 1$, so that one could neglect the effect of $\omega_c$ as a first approximation. Nevertheless, to work exactly at the sweet spot, we need to shift the frequency of the modulated transmon by a fixed amount $\omega_c$ by  fabricating the transmons with a frequency  $\omega\to\omega-\omega_c$; in this way, the dc component of the modulation compensates the fixed frequency shift and effectively no frequency difference appears between the modulated and the non-modulated transmons. As suggested in the main text, by constructing a subset of transmons with a slightly different frequency, it is possible to find the optimal values of $T$ and $\tau$ that compensate the shift a posteriori.

\begin{figure*}[t]
	\centering
	\includegraphics[width=1.0\linewidth]{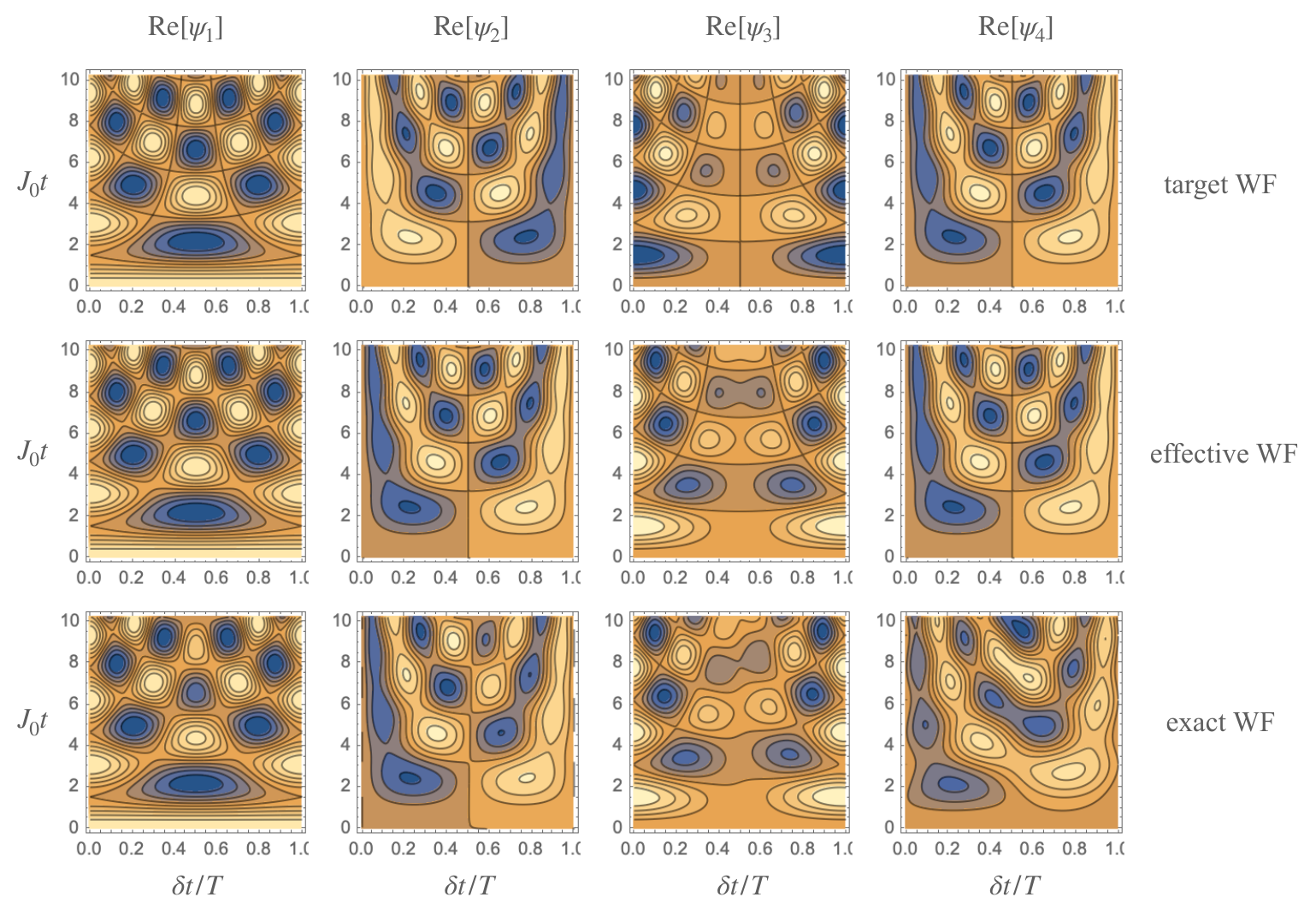}
	\caption{Evolution in time of the real part of the wave function components,  ${\rm Re}~\psi_j(t)$ with $j=1,\ldots,N_s$, for a $N_s=4$ transmons chain in the $N_p=1$ particle sector for three different conditions: (top panels) the Hamiltonian is the target one Eq.~\eqref{Eq:Htarget} and the time-evolution is for different values of the phase $\phi/2\pi=\delta t/T$, (middle panels) the Hamiltonian is the effective one in the fundamental Floquet band, and (bottom panels) the Hamiltonian is the exact one Eq.~\eqref{Eq:Hmodulated}, all shown as a function of the time-shift $\delta t/T$. The parameters are $J_0=0.041~{\rm GHz}$, $T=0.5~{\rm ns}$, $\tau/T=0.3$, and $J_0'=J_0 e^{2\pi\tau/T}$.}
	\label{SM-Fig1}
\end{figure*}

\subsection{Source signal}

Levitons are Lorentzian pulses that can be generated through a proper series of harmonic pulses. We subtract a constant from the modulation, in a way that the minimum of the modulation is zero, as in the case of working at the sweet spot,
\begin{eqnarray}
    \delta\omega(t)&=&\sum_k\frac{2\tau}{(t-kT)^2+\tau^2}-\frac{2\pi}{T}(1+c)\nonumber\\
    &=&\frac{2\pi}{T}\left[2\sum_{n=1}^\infty e^{-2\pi n\tau/T} \cos(2\pi nt/T)-c\right]\nonumber\\
    &=&\frac{2\pi}{T}\left[\frac{\kappa z}{1-\kappa z}+\frac{\kappa}{z-\kappa}-c\right]
\end{eqnarray}
with $c=\tanh(\pi\tau/T)-1$, $z=e^{2\pi it/T}$, and $\kappa=e^{-2\pi \tau/T}$.

Considering only the first term (the quadratic one) of the expansion of $\delta\omega$ in terms of $\varphi_x$ we obtain
\begin{equation}
    \varphi_x(t)=4\sqrt{\frac{2\pi}{T\omega}}\sqrt{\frac{\kappa z}{1-\kappa z}+\frac{\kappa}{z-\kappa}-c},
\end{equation}
that can be generated with a series of pulses 
\begin{equation}
    \varphi_x(t)=a_0+2\sum_{l=1}^\infty a_l\cos(2\pi l t/T),
\end{equation}
whose Fourier components are
\begin{eqnarray}
    a_l&=& 4\sqrt{\frac{2}{\pi T\omega}}\int_{-1}^\kappa dx x^{l-1}\sqrt{\frac{x}{\kappa-x}+\frac{1}{x\kappa-1}-c},~~~~~\\
    a_0&=&-2\sum_{l=1}^{\infty}(-1)^la_l.
\end{eqnarray}

\subsection{Full time-dependent dynamics}

In order to check the prediction of the Floquet theory we check the result for the effective model in the fundamental Floquet band and the exact dynamics generated by the time-dependent modulation against the target evolution with constant uniform hopping and a flux $\phi$. In order to keep the analysis simple, we restrict to the single-particle spectrum and solve the dynamics for a set of $N_s=4$ transmon. In this subspace, the target Hamiltonian in the frame rotating at frequency $\omega$ reads
\begin{equation}\label{Eq:Htarget}
H_\phi=-J_0\left(\begin{array}{ccccc}
0 & 1 & 0 & 1\\
1 & 0 & e^{i\phi} & 0\\
0 & e^{-i\phi} & 0 & 1\\
1 & 0 & 1 & 0
\end{array}
\right),
\end{equation} 
where we have gauged the dependence on the external flux between the second and third site \cite{shastry1990,osterloh2000}. In turn, the evolution in the fundamental Floquet band is described by the effective Hamiltonian 
\begin{equation}\label{Eq:Heffective}
H_{\rm eff}=\left(\begin{array}{ccccc}
0 & -J_0'{\cal I}_2 & 0 & -J_0\\
-J_0'{\cal I}_2 & 0 & -J_0{\cal I}_1 & 0\\
0 & -J_0{\cal I}^*_1 & 0 & -J_0'{\cal I}_2\\
-J_0 & 0 & -J_0'{\cal I}_2 & 0
\end{array}
\right),
\end{equation}
where ${\cal I}_1$ and ${\cal I}_2$ are given by Eq.~\eqref{Eq:I1} and  Eq.~\eqref{Eq:I2}, respectively, and we recall that ${\cal I}_2$ is negative and real, whereas ${\cal I}_1$ is complex. Finally, the time-dependent Hamiltonian with the modulation $\delta\omega(t)$ gauged to the hopping between the transmons  acquires the form 
\begin{equation}\label{Eq:Hmodulated}
H(t)=\left(\begin{array}{cccc}
0 & -J_0' & 0 & -J_0\\
-J_0' & -\delta\omega(t) & -J_0 & 0 \\
0 & -J_0 & -\delta\omega(t-\delta t) & -J_0'\\
-J_0 & 0 & -J_0' & 0 
\end{array}
\right).
\end{equation} 
For practical reasons, we numerically simulate the dynamics after the unitary transformation $U(t)$ given in the main text, that moves the time-dependence to the hoppings.  
We set as initial state an excitation in the site 1, $|\psi(t=0)\rangle=|1,0,0,0\rangle$,  and monitor the evolution of the wave function as a function of time $t$ and time-shift $\delta t$. The real parts of the different components of the wave functions are shown in Fig.~\ref{SM-Fig1} (top panels) for the time-evolution of under the target Hamiltonian Eq.~\eqref{Eq:Htarget}, that serves as reference to check the effect of the modulation,  in Fig.~\ref{SM-Fig1} (middle panels) for  the evolution under the effective Hamiltonian Eq.~\eqref{Eq:Heffective}, and in Fig.~\ref{SM-Fig1} (bottom panels) for the evolution under the modulated Hamiltonian Eq.~\eqref{Eq:Hmodulated}, calculated numerically. We find that the optimal values of the Lorentzian width are $\tau/T\sim 0.2\div 0.4$. This is because the rescaling of $J_0'$ increases the coupling to higher Floquet bands for values $\tau/T\gtrsim 0.4$, and for values $\tau/T\lesssim 0.2$ we have strong dependence of the hopping rate $J$ on the time-shift $\delta t$. Indeed, as shown in the previous sections, for $J_0=0.041~{\rm GHz}$ and $\tau/T=0.3$, in order to avoid the coupling to the first Floquet band, it is necessary that $T\ll 0.5~{\rm ns}$. This can be appreciated in the bottom panels of Fig.~\ref{SM-Fig1}, where a slight asymmetry in the components of the exact wave function under the $\delta t\to T-\delta t$ transformation arise from coupling to the first Floquet bands.

\begin{figure}[t]
	\centering
	\includegraphics[width=1.0\linewidth]{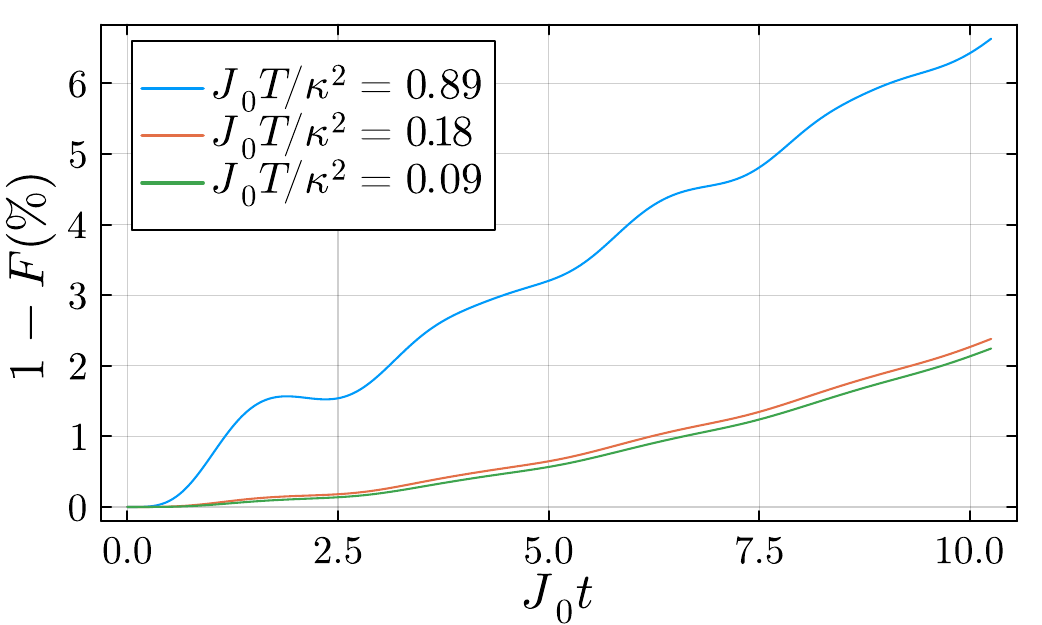}
	\caption{Infidelity of the time-dependent exact evolution in the $N_p=1$ sector of the Floquet modulated Hamiltonian with respect to the target evolution. Parameters are $J_0=0.041~{\rm GHz}$, $\tau/T=0.3$, and we vary $T$.}
	\label{SM-Fig2}
\end{figure}

In order to test the goodness of the Floquet modulation, we also study the fidelity $F=|\langle\psi_{\rm target}|P|\psi_{\rm exact}\rangle|$ integrated over the time/phase-shift as a function of time for different values of $T$, while keeping constant ratio $\tau/T=0.3$. The matrix $P={\rm diag}(1,-1,-1,1)$ accounts for the $\pi$ phase shifts due to the Levitons in the external links, as in Eq.~\eqref{Eq:I2}. As expected the larger the Floquet frequency $2\pi/T$ the better is the fidelity, as shown in Fig.~\eqref{SM-Fig2} where we plot the infidelity $1-F$ for four different values of $T$. For $T=1~{\rm ns}$ we see that the sampling at half period shows oscillations at the Floquet frequency. For higher values of the Floquet frequency we see that the error accumulated grows to few percents after 10 periods of the coherent oscillations set by $J_0$.

We point out that in order for the protocol to be experimentally feasible, the amplitude of the Levitons has to be a fraction of the bare transmon frequency $\omega$. In particular, as pointed out in the previous sections, the regime of compatibility is $2\pi/T<2\pi/\tau<\omega$. For values of $T=0.5~{\rm ns}$ and $\tau/T=0.3$, the bare frequency needs to be $\omega\gtrsim 8~{\rm GHz}$, which is slightly incompatible with the transmon frequency $\omega=3.9~{\rm GHz}$ in the experiment of Ref.~\cite{fedorov2021photon}. Nevertheless, as pointed out in the main text, the many-body evolution depends on the values of $J$, $U$, and $\omega_d-\omega$, so that an increase of $\omega$ is not incompatible with the results of the many-body simulations. What matters to be able to observe the soliton bands and their fractional periodicity is the ratio $J/U$. 

The values chosen for $U$ and $J$ have a good ratio for the emergence of soliton bands that are well detached from the single-particle band. It follows that it is necessary to increase $\omega$ while keeping $J$ and $U$ constant, or decrease $J$ and $U$ while keeping their ratio constant. This can be achieved through a rescaling 
\begin{equation}
E_J\to \lambda E_J, \qquad E_C\to E_C/\lambda, \qquad C_0\to C_0/\lambda, 
\end{equation}
with $\lambda>1$, such that we have 
\begin{eqnarray}
\omega&\to& \omega+E_C(1-1/\lambda)>\omega,\nonumber\\
U&\to& U/\lambda<U,\nonumber\\
J_0&\to& J_0/\lambda<J_0, 
\end{eqnarray}
while keeping the ratio $J_0/U$ constant.

\begin{figure*}[t]
	\centering
	\includegraphics[width=1.0\linewidth]{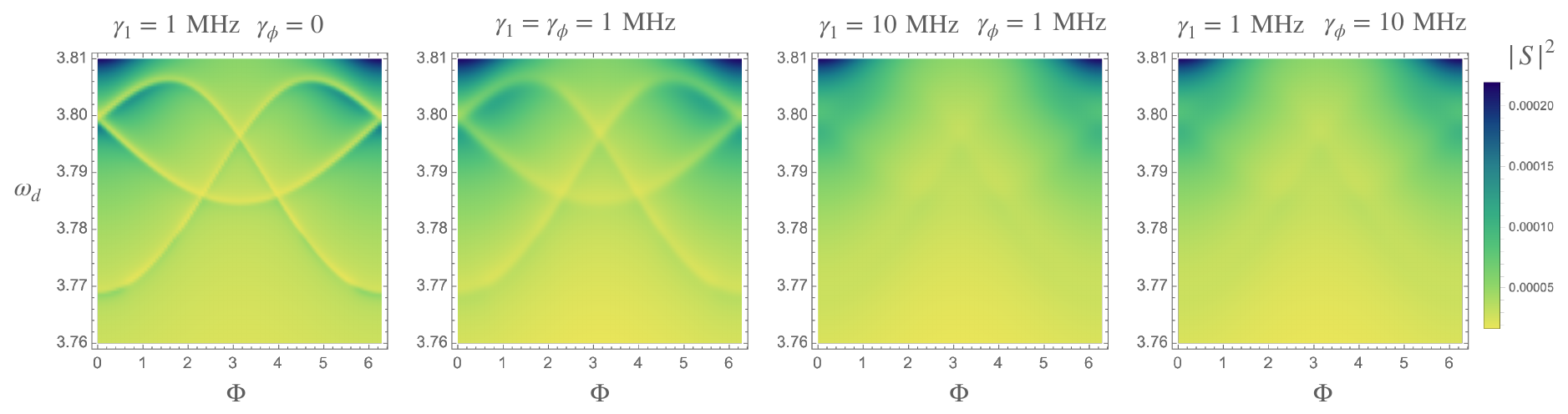}
	\caption{Two-photon spectrum for a chain of $N_s=3$ sites,  as a function of the flux $\Phi$ and the driving frequency $\omega_d$, for  $\Omega=0.01~{\rm GHz}$ and $\Gamma=1~{\rm MHz}$, $\omega=3.9~{\rm GHz}$, and $U=0.188~{\rm GHz}$, for different values of the transmon  relaxation and dephasing rates.}
	\label{SM-Fig3}
\end{figure*}

\begin{table}[b]
\caption{Optimal choice of energy scales (in GHz), partly from Ref.~\cite{fedorov2021photon}.}
\centering
\begin{tabular}{c c c c c c c c c }
\hline\hline
$\gamma_\phi$ & $\gamma_1$ & $\Gamma$ & $J_0$ & $\Omega$ & U & $1/T$ & $1/\tau$ & $\omega$ \\ [0.5ex] 
\hline
$<$0.001 & 0.001 & 0.01 & 0.041 & 0.04 & 0.188 & 2.0 & 6.67 & 8.0 \\
 [1ex]
\hline
\end{tabular}
\label{table:nonlin}
\end{table}

\section{Perturbative approach to the many-body spectrum}

Here, we provide details of the many-body spectrum of the Bose-Hubbard model. In particular, we focus on soliton/stack-like states. Following \cite{mansikkamaki2022beyond}, for large $U$, the $N_p$-bright soliton approximately corresponds to stack-like Fock state  $|N_p\rangle_j$. The hopping amplitude to displace such  state can be written as \begin{equation}
T_{(N_p)}={}_{j+1}\langle N_p|_{j}\langle 0|H_t(G_0(\epsilon)H_t)^{N_p-1}|N_p\rangle_j|0\rangle_{j+1}, 
\end{equation}
with $H_t=-J e^{i\Phi/N_s}a^\dag_{j+1}a_j$, and the unperturbed Green's function 
\begin{equation}
G^{-1}_0(\epsilon)=\epsilon-\sum_j(\omega n_j+Un_j(n_j-1)/2), 
\end{equation}
where $\epsilon=N_p\omega-UN_p(N_p-1)/2$ is the unperturbed energy of the $N_p$-stack states. A direct calculation show that the complex amplitude of the $N_p$ soliton/stack-like state reads
\begin{equation}
    T_{(N_p)}=-J_{(N_p)}e^{iN_p\Phi/N_s}
\end{equation}
with $J_{(N_p)}=\frac{J N_p}{(N_p-1)!}(J/U)^{N_p-1}$ in agreement with the result of Ref. 
\cite{mansikkamaki2022beyond}. This allows us to directly compute the dispersion of the $N_p$ soliton band as provided in the main text.

Stack-like states allow us also to asses the matrix elements between different propagating states via second order perturbation theory. Defining 
\begin{equation}
|\tilde{N}_p\rangle_k=\frac{1}{\sqrt{N_s}}\sum_{j=1}^{N_s}e^{i (2\pi jk+N_p\Phi)/N_s}|N_p\rangle_j, 
\end{equation}
the matrix element between different solitons is found to be
\begin{eqnarray}
    \Delta_{k,k'}&=&
    \frac{\Omega^2}{4}{}_k\langle\tilde{N}_p|(a_1+a_1^\dag)G_0(\epsilon)(a_1+a_1^\dag)|\tilde{N}_p\rangle_{k'}\nonumber\\
    &=&-\frac{\Omega^2(\omega+U)e^{2\pi i(k-k')/N_s}}{4N_s(\omega-U(N_p-1))(\omega-U N_p)}.
\end{eqnarray}
Estimates of the gaps that open at the crossing points in the soliton states are provided in the main text and they are one order of magnitude smaller than those resulting from the numerical simulation. We ascribe this features to the driven-dissipative dynamics in presence of interactions, that will be discussed in the next sections.

\section{Master equation numerical approach}

The Lindbladian constitutes a completely positive, trace-preserving linear map between density operators. It can be split in two terms, the first one can be described through a non-Hermitian Hamiltonian $K$,
\begin{equation}
K=H-\frac{i}{2}\sum_{j=1}^N(\gamma_1n_j+\gamma_\phi n_j^2).
\end{equation}
Its action on the density matrix can be written as
\begin{equation}
{\cal L}_{nh}[\hat{\rho}]=-i(K\hat{\rho}-\hat{\rho}K^\dag),
\end{equation}
and introduces a decaying part to the time-evolution of the matrix elements of the density matrix. The second part of the dissipator is given by,
\begin{equation}
{\cal L}_{tp}[\hat{\rho}]=\sum_{j=1}^N \gamma_1 a_j\hat{\rho}a^\dag_j+\gamma_\phi n_j\hat{\rho}n_j,
\end{equation}
and it contributes to preserve the completely positive, trace-1 character of the density matrix, so that its role cannot be simply neglected. 

\begin{figure*}[t]
	\centering
	\includegraphics[width=0.8\linewidth]{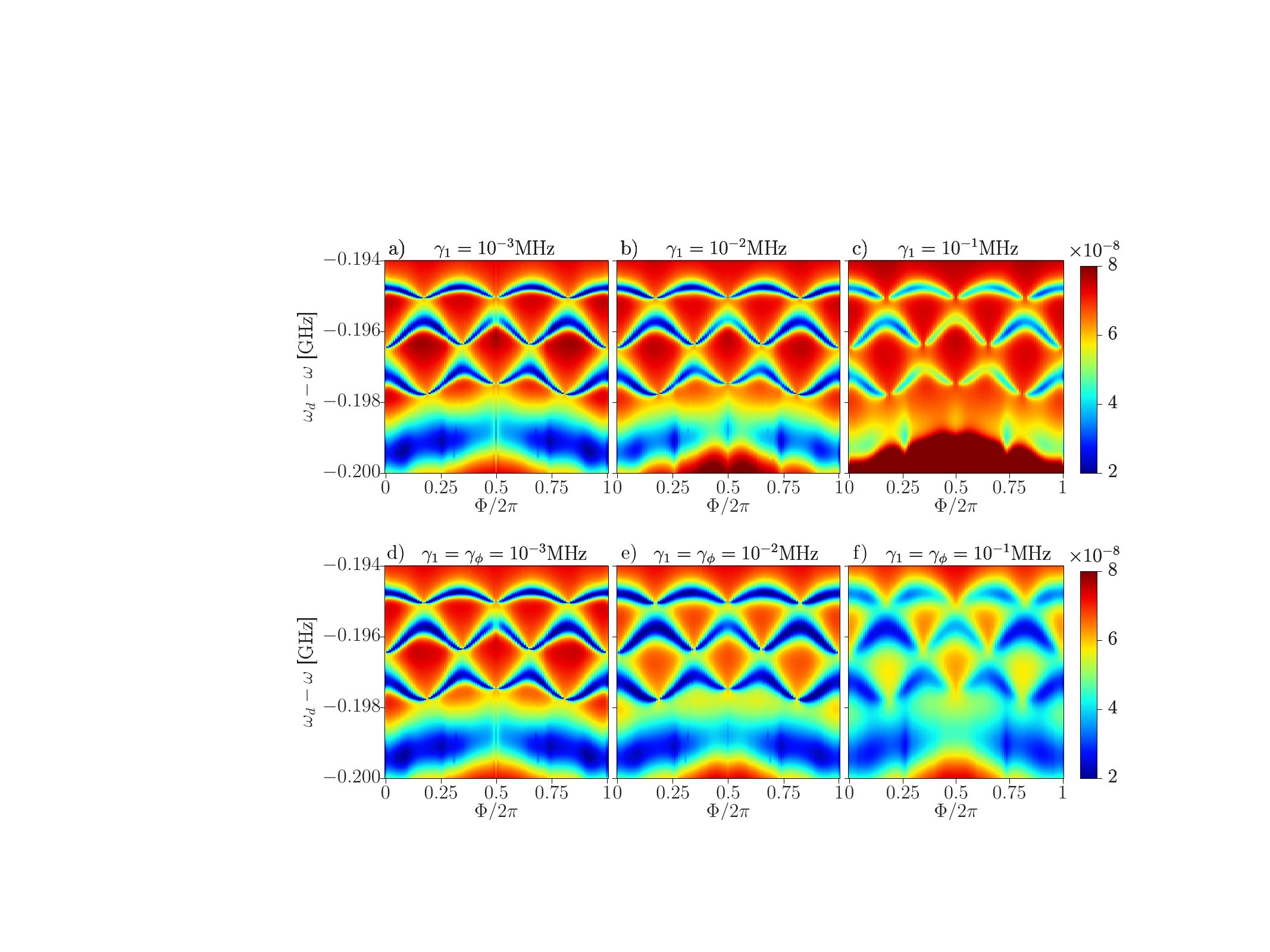}
	\caption{Three-photon spectrum for a chain of $N_s=4$ sites,  as a function of the flux $\Phi$ and the driving frequency $\omega_d$, for $J=0.041~{\rm GHz}$,  $\Omega=0.04~{\rm GHz}$ and $\Gamma=0.1~{\rm MHz}$, $\omega=3.9~{\rm GHz}$, and $U=0.188~{\rm GHz}$, for different values of the transmon  relaxation rate $\gamma_1$. (a,b,c)  neglecting dephasing, $\gamma_\phi=0$; (d,e,f) $\gamma_\phi=\gamma_1$.}
	\label{SM-Fig4}
\end{figure*}

A linear map on the density matrix can be expressed in a matrix form by vectorializing the density matrix as
\begin{equation}
\hat{\rho}=\sum_{n,m}\rho_{n,m}|n\rangle\langle m|\equiv 
\sum_{n,m}\rho_{n,\tilde{m}}|n\rangle| \tilde{m}\rangle,
\end{equation}
where $\{|n\rangle\}$ is a basis of the Hilbert space of the Hamiltonian $H$. The procedure of vectorialization then amounts to a tensor product of the original Hilbert space states with itself, that clearly squares its dimension. Given an analogous matrix representation of the non-Hermitian Hamiltonian in a basis of states $\{|n\rangle\}$, $K=\sum_{n,m}K_{n,\tilde{m}}|n\rangle |\tilde{m}\rangle$, the element $(n,\tilde{m})$ of the map can be written as 
\begin{eqnarray}
i{\cal L}_{nh}[\hat\rho]_{n,\tilde{m}}&=&(K\hat{\rho}-\hat{\rho}K^\dag)_{n,\tilde{m}}\nonumber\\
&=&\sum_l K_{n,\tilde{l}}\rho_{l,\tilde{m}}-K^*_{\tilde{m},l}\rho_{n,\tilde{l}}.
\end{eqnarray}
Analogously, a matrix representation of the second part of the Lindbladian can be obtained starting from a matrix representation of the operators involved in ${\cal L}_{tp}[\hat\rho]=\sum_{p}\hat{O}_p\hat\rho \hat{O}_p^\dag$. By expressing $\hat{O}_p=\sum_{n,m}O_{n,\tilde{m}}|n\rangle|\tilde{m}\rangle$, we have that 
\begin{equation}
{\cal L}_{tp}[\hat\rho]_{n,\tilde{m}}=\sum_{p}\sum_{k,l}(O_p)_{n,\tilde{k}}(O^\dag_p)_{l,\tilde{m}}\rho_{k,\tilde{l}}.
\end{equation}
For the numerical simulations we work with a Fock state representation of the boson states. The Lindbladian ${\cal{L}}$ can be cast in a matrix form by vectorializing the density matrix and the steady state is given by the eigenstate corresponding to the zero eigenvalue, within numerical precision. The dimension $L$ of the Hilbert space and hence size of the matrix representation of the Hamiltonian $L\times L$ grows quickly with the number of sites $N_s$ and maximum number of excitations $N_p$, $L=\sum_{n=0}^{N_p}\binom{n+N_s-1}{n}$. Correspondingly, the size of the Lindbladian matrix grows as $L^2\times L^2$.

 \begin{figure}[b]
	\centering
	\includegraphics[width=1.0\linewidth]{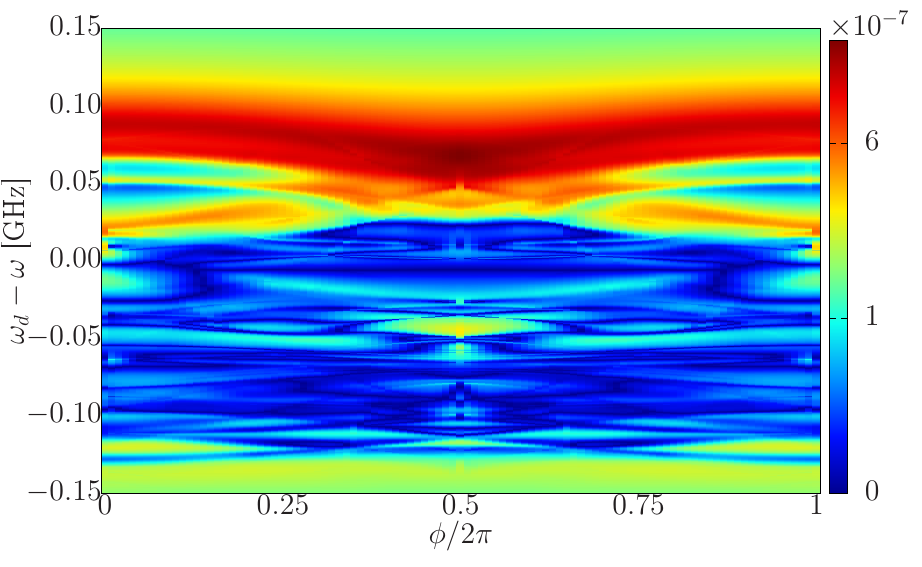}
	\caption{ Spectrum $|S(\omega_d)|^2$ as a function of the driving frequency $\omega_d$ and the flux $\Phi$ for $\Omega=0.04~{\rm GHz}$ in the region frequency around the single-photon and two-photon resonances. The other parameters are as given in the main text and the color bar is made non-linear by setting $z(x,y)^{2/5}$. Note that the range of values covered by the color scale is much smaller than in Fig.~3 of the main text (see also Fig.~\ref{SM-Fig5}).}
	\label{SM-Fig5}
\end{figure}

To observe flux fractionalization in the $N_p=2$ photon sectors we assume low transmon relaxation rate, $\gamma_1=1~{\rm MHz}$, low coupling rate to the transmission line resonator,  $\Gamma=1\div 10~ {\rm MHz}$ and zero transmon pure dephasing rate, $\gamma_\phi=0$. As pointed out in the main text, we expect the oscillations in the $N_p$ sector soliton to be visible as long as $N_p \gamma_1,\gamma_\phi\ll J_{(N_p)}/N_s$. In Fig.~\ref{SM-Fig3} we show the fading of the oscillations as we increase the relaxation and dephasing rates. The effective hopping for the $N_p=2$ soliton with $J=0.041~{\rm GHz}$ and $U=0.188~{\rm GHz}$ is $J_{(2)}=0.018~{\rm GHz}$, so that in the left panels, where $J_{(2)}/3> 2\gamma_1,\gamma_\phi$ oscillations are visible, whereas in the right panels, where $J_{(2)}/3< 2\gamma_1,\gamma_\phi$ oscillations are lost.

In the main text we show the results of the simulations for different driving strength $\Omega$ and for $\gamma_1=10^{-3}~{\rm MHz}$. This value of the relaxation rate is on the order of the best relaxation rates recorded for superconducting qubits \cite{milul2023,ganjam2024}. In Fig.~\ref{SM-Fig4}, we complement the description by  
studying the fading of the oscillations with flux in the $N_p=3$ sector by increasing the decoherence rates. In the top panel we set $\gamma_\phi=0$ and study the effect of  $\gamma_1$ only: we clearly see that oscillations persist up to $\gamma_1=0.1~{\rm MHz}$, corresponding to a $T_1$ on order of 10 $\mu$s, that is well below the current relaxation times. Indeed, for $N_p=3$ and $N_s=4$ the upper bound for the relaxation rate is $\gamma_1<J_{(3)}/12=0.24~{\rm MHz}$, well within state-of-the-art relaxation rates. We point out that the decoherence rate is dominated by the relaxation rate, that scales proportionally with the number of photons. In turn, there is no clear indication that the dephasing rate scales with the number of excitations. Nevertheless, we studied the impact of a finite dephasing rate $\gamma_\phi=\gamma_1$. The result is shown in the bottom panels of Fig.~\ref{SM-Fig4}: for $\gamma_\phi=\gamma_1=0.001 - 0.01$ MHz the oscillations remain clearly visible, whereas for the case of $\gamma_\phi=\gamma_1=0.1$ MHz a more severe broadening is observed. We conclude that the observation of the oscillations in the $N_p=3$ sector for a chain of $N_s=4$ transmons is within the current experimental possibilities.

In addition, in Fig.~\ref{SM-Fig5}, we show the plot of the signal for driving $\Omega=0.04~{\rm GHz}$ in the region around the single-photon peak. We see a signal that is strongly modified with respect to the weak driving one. In Fig.~\ref{SM-Fig6}, top panel, we plot the output signal $S(\omega_d)$ as a function of the driving frequency $\omega_d$ and the driving strength $\Omega$ at zero flux, with specific cuts at different $\Omega$ in the bottom panel. We clearly see that the signal for increased driving strength $\Omega$ is highly modified from the weak driving case. In particular, the single-photon peak splits and transfers its spectral weight to higher photon resonances, that are located at the expected many-body energies. We then see that the two-photon and three-photon peaks are very narrow resonances for weak relaxation rate $\Gamma$, surrounded by larger dips. The peaks are washed out by increasing $\Gamma$, and only dips remain, that nevertheless show the expected evolution with the unperturbed many-body energies, as shown in the main text. In the next section we study the single site driven-dissipative dynamics in presence of Hubbard interaction and ascribe to the combination of driving, dissipation, and interaction the complicate spectrum evolution.

\begin{figure}[t]
	\centering
	\includegraphics[width=1.0\linewidth]{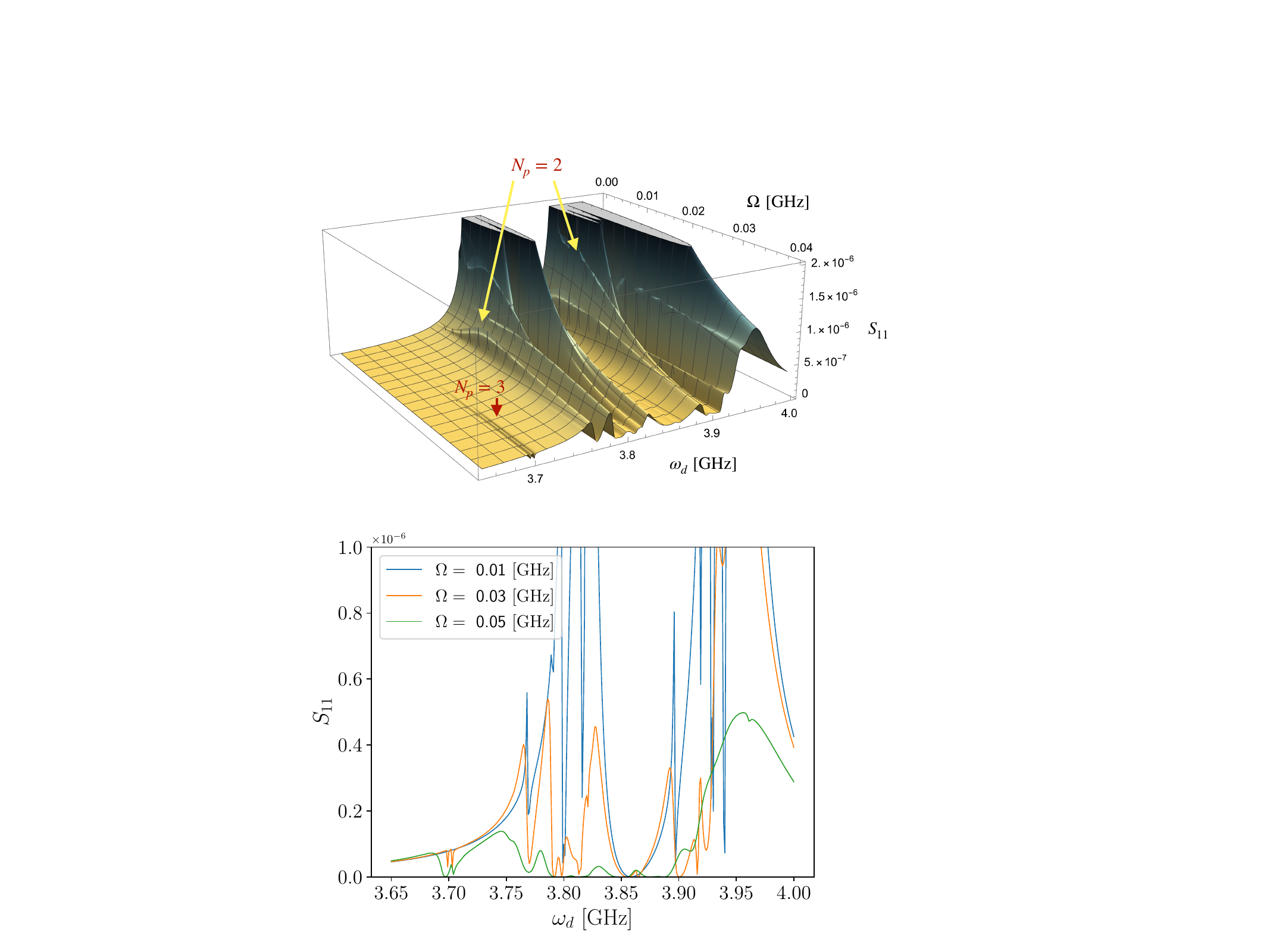}
	\caption{Top: reflection spectrum as a function of the driving frequency $\omega_d$ and the driving amplitude $\Omega$, for $\omega=3.9$ GHz, $U=0.188~{\rm GHz}$, $J=0.041~{\rm GHz}$, $\Gamma=0.1~{\rm MHz}$, $\gamma_1=0.001~{\rm MHz}$, and we neglect $\gamma_\phi$. \label{SM-Fig6}}
\end{figure}

\section{Single site driven-dissipative dynamics}

Here, we study the problem of a single site boson with interaction, coherent driving and dissipation through the transmission-line resonator. The time evolution of the density operator is governed by the Lindbladian
\begin{equation}
    \frac{\partial \rho}{\partial t}=-i[H,\rho]+\Gamma\left(a\rho a^\dag-\frac{1}{2}(a^\dag a\rho+{\rm H.c.})\right),
\end{equation}
with the Hamiltonian
\begin{equation}
    H=\omega a^\dag a-\frac{U}{2}a^\dag a(a^\dag a-1)+\Omega(t)(a+a^\dag),
\end{equation}
with $\Omega(t)=\Omega\cos(\omega_dt)$. 
We then move to a frame rotating at frequency $\omega_d$ and neglect counter rotating terms. This yields a time independent Hamiltonian, 
\begin{equation}
    H_{\rm rwa}=(\omega-\omega_d) a^\dag a-\frac{U}{2}a^\dag a(a^\dag a-1)+\frac{\Omega}{2}(a+a^\dag).
\end{equation}
In the long time the system approaches a steady state that solves $\partial \rho_{ss}/\partial t=0$ and the associated signal is  given by $|S(\omega_d)|^2$ where
\begin{equation}
    S(\omega_d)=\frac{i\Gamma}{\Omega}{\rm Tr}[a^\dag \rho_{ss}].
\end{equation}
The problem of a driven-dissipative dynamics in presence of interaction has been studied in the literature and the first exact solution  has been provided by Drummond and Walls \cite{drummond1980quantum} by means of the generalized $P$-representation of the density matrix in terms of coherent states \cite{drummond1980generalised}. A closed form for the Wigner function has been provided by Kheruntsyan in presence of also two-photon losses \cite{kheruntsyan1999wigner}, and successive extension to the case of two-photon driving and two-photon losses have been provided through the generalized $P$-representation \cite{bartolo2016exact}, and through the coherent quantum-absorber method relying on exploiting the Segal-Bargmann representation of Fock space \cite{stannigel2012driven-dissipative,roberts2020newexact}

\begin{figure}[t]
	\centering
	\includegraphics[width=1.0\linewidth]{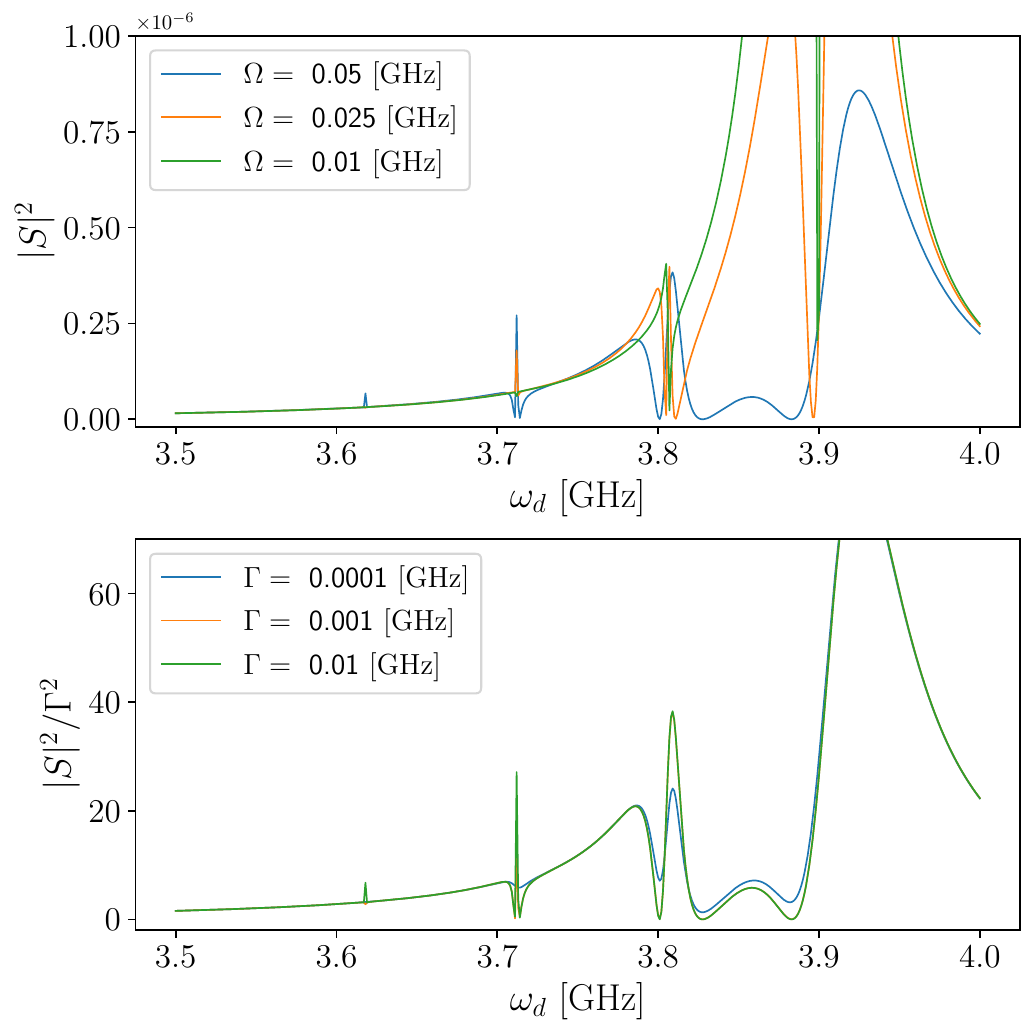}
	\caption{Modulus squared of the signal $|S(\omega_d)|^2$ as a function of the driving frequency $\omega_d$. Top panel: signal for different driving strength $\Omega$ and fixed rate $\Gamma=0.0001~{\rm GHz}$. Bottom panel: signal for different relaxation rates $\Gamma$ and fixed driving strength $\Omega=0.05~{\rm GHz}$.}
	\label{SM-Fig6}
\end{figure}

In the complex-$P$ representation the density matrix is written as
\begin{equation}
    \hat{\rho}=\int_{{\cal C}}d\alpha\int_{{\cal C}'}d\beta \frac{|\alpha\rangle\langle\beta^*|}{\langle\beta^*|\alpha\rangle}P(\alpha,\beta)
\end{equation}
where the closed integration contours ${\cal C}$ and ${\cal C}'$ must be carefully chosen to encircle all the singularities of the function $P(\alpha,\beta)$. The Lindbald equation for the density matrix then reads
\begin{eqnarray}
    \frac{\partial P(\alpha,\beta)}{\partial t}&=&\left[\frac{\partial}{\partial\alpha}(\bar{\gamma}\alpha+2\chi \alpha^2\beta+i\Omega/2)-\chi\frac{\partial^2}{\partial \alpha^2} \alpha^2\right.\nonumber\\
    &+&\left.\frac{\partial}{\partial \beta}(\bar{\gamma}^*\beta+2\chi^*\beta^2\alpha-i\Omega/2)-\chi^*\frac{\partial^2}{\partial \beta^2} \beta^2\right]\nonumber\\
    &\times&P(\alpha,\beta)
\end{eqnarray}
where $\bar{\gamma}=i\omega+\Gamma/2$ and $\chi=-iU/2$, and the steady state solution reads \cite{drummond1980quantum}
\begin{equation}
    P_{ss}(\alpha,\beta)=\frac{\alpha^{\lambda-2}\beta^{\lambda^*-2}}{\cal N}\exp\left(2\alpha\beta+\frac{\epsilon}{\alpha}+\frac{\epsilon^*}{\beta}\right),
\end{equation}
with $\epsilon=-i\Omega/(2\chi)$, $\lambda=\bar{\gamma}/\chi$, and ${\cal N}$ a normalization constant. The latter, together with the general unnormalized expression for the moments of the field are expressed through the integral \cite{drummond1980quantum}
\begin{eqnarray}
    {\cal I}(a,b,\epsilon)&=&\int_{{\cal C}}d\alpha\int_{{\cal C}'}d\beta \alpha^{c-2}\beta^{d-2}e^{\frac{\epsilon}{\alpha}+\frac{\epsilon^*}{\beta}+2\alpha \beta}\nonumber\\
    &=&-4\pi^2\frac{\epsilon^{c-1}(\epsilon^*)^{d-1}}{\Gamma(c)\Gamma(d)}{_0F_2(a,b,2|\epsilon|^2)}
\end{eqnarray}
in terms of the function
\begin{equation}
    {}_0F_2(a,b,z)=\sum_{k=0}^\infty\frac{z^k\Gamma(a)\Gamma(b)}{k!\Gamma(k+a)\Gamma(k+b)},
\end{equation}
so that ${\cal N}={\cal I}(\lambda,\lambda^*)$ and the general expression for the moments reads \cite{drummond1980quantum}
\begin{eqnarray}
    \langle (a^\dag)^ka^l\rangle&=&\frac{{\cal I}(\lambda+l,\lambda^*+k)}{{\cal I}(\lambda,\lambda^*)}
\end{eqnarray}
It follows that the desired output field is expressed as
\begin{eqnarray}\label{EqSM:output}
    \langle a^\dag\rangle&=&\frac{-\Omega/2}{\omega+i\Gamma/2}\frac{{}_0F_2\left(\frac{-\omega+i\Gamma/2}{U},\frac{-\omega-i\Gamma/2}{U}+1,\Omega^2/(2U^2)\right)}{{}_0F_2\left(\frac{-\omega+i\Gamma/2}{U},\frac{-\omega-i\Gamma/2}{U},\Omega^2/(2U^2)\right)}.\nonumber\\
\end{eqnarray}
We see that in the limit for $U\to 0$ the output field asymptotically approaches the coherent state solution $\langle a^\dag\rangle_0=-\Omega/(\omega+i\Gamma/2)$ of a driven-dissipative harmonic oscillator. 

In Fig.~\ref{SM-Fig6} we plot the output signal as function of the driving strength $\Omega$ and the relaxation rate $\Gamma$. In the top panel we see that as we increase the driving strength the main single-particle peak splits and other peaks at the two-photon and three-photon resonances appear. As the spectral weight is gradually transferred from the single-photon peak to the other resonances a complicate response featuring dips and splittings appears, that is totally due to the combination of driving and interaction. In the bottom panel we show the evolution of the output signal by increasing the relaxation rate. A general intuitive trend appears, in which fine structure details of the signal are washed out by increasing $\Gamma$. 

The trend revealed by the exact solution for the single site problem captures the more complicated one associated to the chain of bosons and shown in Fig.~\ref{SM-Fig5}. We then conclude that the non-trivial evolution of the signal for increased driving is associated to the single-site driven dissipative dynamics with interactions.

As a check, we have confirmed that the results of Eq.~(\ref{EqSM:output}) and the results of the numerical calculation of the steady state for a single site driven-dissipative Bose-Hubbard model coincide.

\end{document}